\documentclass{article}

\usepackage{PRIMEarxiv}

\usepackage[utf8]{inputenc} 
\usepackage[T1]{fontenc}    
\usepackage{hyperref}       
\usepackage{xurl}            
\usepackage{booktabs}       
\usepackage{amsfonts}       
\usepackage{nicefrac}       
\usepackage{microtype}      
\usepackage{lipsum}
\usepackage{fancyhdr}       
\usepackage{graphicx}       
\graphicspath{{media/}}     
\usepackage{multirow}
\usepackage{fontawesome}
\title{Towards a Responsible AI Metrics Catalogue: A Collection of Metrics for AI Accountability}

\author{
 Boming Xia\textsuperscript{1,2}, Qinghua Lu\textsuperscript{1,2}, Liming Zhu\textsuperscript{1,2}, Sung Une (Sunny) Lee\textsuperscript{1}, Yue Liu\textsuperscript{1,2}, and Zhenchang Xing\textsuperscript{1,3} \\
  \textsuperscript{1}CSIRO's Data61, \textsuperscript{2}University of New South Wales, \textsuperscript{3}Australian National University\\
\texttt{firstname.lastname@data61.csiro.au} 
}

\begin{document}
\maketitle

\begin{abstract}
Artificial Intelligence (AI), particularly through the advent of large-scale generative AI (GenAI) models such as Large Language Models (LLMs), has become a transformative element in contemporary technology. While these models have unlocked new possibilities, they simultaneously present significant challenges, such as concerns over data privacy and the propensity to generate misleading or fabricated content. 
Current frameworks for Responsible AI (RAI) often fall short in providing the granular guidance necessary for tangible application, especially for \textit{Accountability}—a principle that is pivotal for ensuring transparent and auditable decision-making, bolstering public trust, and meeting increasing regulatory expectations.
This study bridges the \textit{Accountability gap} by introducing our effort towards a comprehensive metrics catalogue, formulated through a systematic multivocal literature review (MLR) that integrates findings from both academic and grey literature. Our catalogue delineates process metrics that underpin procedural integrity, resource metrics that provide necessary tools and frameworks, and product metrics that  reflect the outputs of AI systems. This tripartite framework is designed to operationalize Accountability in AI, with a special emphasis on addressing the intricacies of GenAI.
\end{abstract}


\section{Introduction}
Artificial Intelligence (AI) serves as the linchpin of contemporary technological innovation, with large-scale generative AI (GenAI) models, exemplified by Large Language Models (LLMs), at the forefront of this paradigm shift \cite{bengio2023managing}. These models have demonstrated remarkable proficiency in generating content across modalities, including text, imagery, and code. However, they also precipitate complex legal, ethical and operational challenges, such as data privacy breaches \cite{zhang2023tag, khan2022subjects}, lack of transparency \cite{bommasani2023foundation}, and the erosion of informational integrity, epitomized by the generation of ``hallucinations''—erroneous or misleading information produced by the models \cite{openai2023gpt4}. These challenges gain further gravitas in critical sectors, where the misuse of AI could lead to biased decision-making and the contentious deployment of dual-use biotechnologies, as underpinned by a recent \textit{US Executive Order on Safe, Secure, and Trustworthy Artificial Intelligence} \cite{WhiteHouse2023AISecurity}.

The imperative for the practical implementation of responsible AI (RAI) is accentuated by the proliferation of high-level principles published by organizations or governments worldwide \cite{RAIMatrix}. Despite providing guidance, these frameworks often lack the granularity necessary for practical implementation, falling short in providing concrete and actionable solutions \cite{xia2023towards}.
In this context, accountability emerges as a pivotal element within RAI \cite{akhgar2022accountability}. It serves as a fundamental principle that for establishing transparent, fair, and ethical AI systems. Through responsibility allocation, accountability enhances governance practices and contributes to the auditability and trustworthiness of AI systems' decision-making processes. The significance of accountability in AI is further underscored by evolving global standards and legislation, such as the proposed EU AI Act \cite{EUAIACT} and the US Blueprint for an AI Bill of Rights \cite{WhiteHouse2023AIBillOfRights}. These legislative developments highlight the necessity for actionable and concrete guidelines that operationalize accountability in AI.

In light of these challenges, our research underscores the pivotal role of a process-oriented approach that encompasses both technical and socio-technical dimensions. As elucidated by Raji et al. \cite{raji2020closing}, procedural justice is predicated on the legitimacy of outcomes derived from equitable and comprehensive processes. This principle is foundational to the development of robust frameworks that enable independent audits, adherence to established standards, and enhanced compliance \cite{lucaj2023ai}. Similarly, the capAI framework \cite{floridi2022capai}, which operationalizes the proposed EU AI Act's directives, emphasizes a process-oriented perspective of AI systems throughout their lifecycle. Our study aligns with this paradigm, concentrating on the development of process metrics as the cornerstone procedural guidelines for the operationalization of AI accountability.

Anchored in a systematic multivocal literature review (MLR) that combines Systematic Literature Review (SLR) and Grey Literature Review (GLR), we aim to transition from aspirational RAI principles and answer the following research question (RQ): \textbf{What are the definitive, actionable metrics for AI risk management with respect to Accountability?} Drawing inspiration from an established software metrics framework \cite{fenton2014software}, we introduce a dedicated AI-centric \textbf{system-level} metrics framework specifically tailored to the principle of Accountability and sensitive to the intricacies of GenAI. This framework delineates metrics into three interrelated categories: \textbf{process metrics}, which establish foundational procedural guidelines; \textbf{resource metrics}, which encompass the necessary tools and frameworks; and \textbf{product metrics}, which include the resultant artifacts. While our current metrics primarily offer a binary (yes/no) evaluation, this work represents an initial step towards developing a more sophisticated metrics catalog. Our approach also extends beyond these binary evaluations to include qualitative considerations of these processes, as depicted in Fig. \ref{Fig:structure}. This comprehensive approach lays the groundwork for an integrated RAI framework that covers all principal RAI principles.
The primary contributions of this paper are as follows:

\begin{figure}[]
  \centering
  \includegraphics[width=0.6\linewidth]{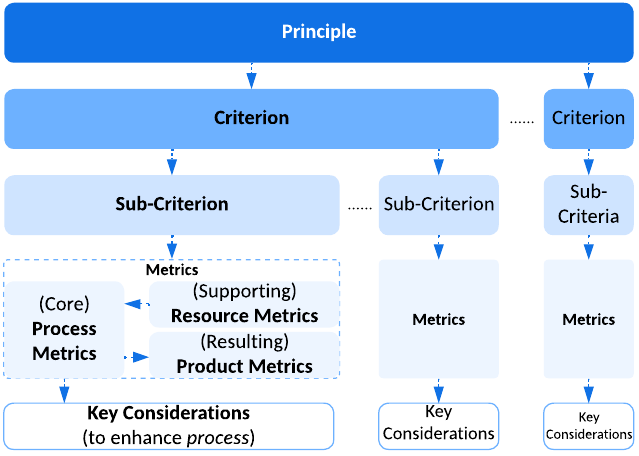}
  \caption{Structure Overview}
  \label{Fig:structure}
\end{figure}

\begin{itemize}
\item \textbf{Development of a Process-centric Metrics Catalogue for AI Accountability}: We introduce a metrics catalogue that operationalizes accountability in AI systems, with a particular focus on GenAI. This catalogue provides essential procedural guidelines for embedding accountability in AI.

\item \textbf{Categorization of AI Accountability Metrics}: We categorize AI accountability metrics into process, resource, and product metrics. This tripartite categorization provides a structured approach to operationalize AI accountability.

\item \textbf{Foundation for a Comprehensive RAI Framework}: This research lays the groundwork for a more expansive framework that integrates all key RAI principle, marking an important step forward in the discourse on RAI practices.
\end{itemize}


\section{Background and Challenges}

\subsection{Accountability}
\subsubsection{The Three Facets of Accountability}
\label{sec:AccConcept}
Accountability, a concept integral to various domains from governance and law to technology and science, is foundational to ethical practice \cite{sinclair1995chameleon, henriksen2021situated}. In the legal realm, accountability is a well-defined term, emphasizing the necessity for individuals or entities to be responsible for their actions and consequences \cite{Acc_Definition}.
Under certain data protection legislation such as the EU's General Data Protection Regulation (GDPR), accountability is articulated as a requirement wherein data controllers need to assume responsibility for compliance and possess demonstrable evidence of their adherence to the regulatory standards and provisions.
Expanding beyond its legal roots, accountability in a broader sense revolves around the answerability \cite{bovens2007analysing, novelli2023accountability}. This principle demands that actors justify their actions to an overseeing authority, which then has the power to impose consequences based on these justifications and the actor's performance.

\begin{figure}[htb]
  \centering
  \includegraphics[width=0.6\linewidth]{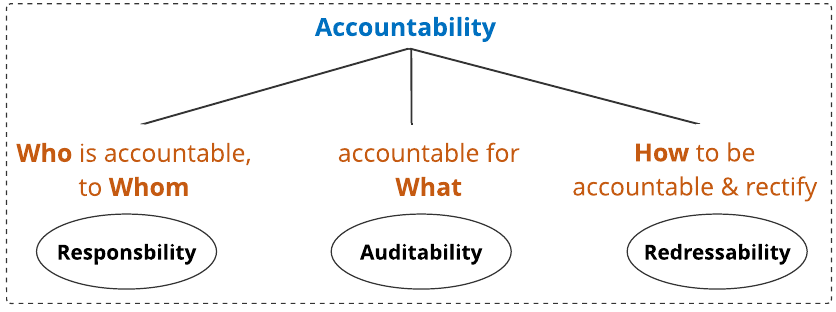}
  \caption{Three Facets of Accountability}
  \label{Fig:AccConcept}
\end{figure}

The broader interpretation of accountability manifests in three interconnected facets: Responsibility, Auditability, and Redressability (see Fig. \ref{Fig:AccConcept}).
\textbf{Responsibility} (Section \ref{Sec:Responsibility}) pertains to the attribution of ownership for actions to individuals or entities by establishing who is accountable to whom within a (cross-)organizational context.
\textbf{Auditability} (Section \ref{Sec:Auditability}) serves as the cornerstone for ``accountable for what.'' It facilitates the systematic assessment of decisions and their outcomes against established criteria. This facet ensures that actions of responsible entities are \textit{demonstrable} with supporting evidence--not only traceable but also defensible. 
\textbf{Redressability} (Section \ref{Sec:Redressability}) addresses the aspect of ``how to be accountable and rectify'' after ascertaining ``who is accountable for what.'' It involves the provision of mechanisms for remediation or compensation when actions lead to adverse outcomes. This facet completes the accountability cycle by providing avenues for remedy and reparation.

These elements ensure that individuals and organizations are not only expected to justify their actions but also that their justifications are backed by evidence and liable to appropriate consequences. This tripartite framework is essential for the ethical operation and maintenance of equitable systems within complex societal structures \cite{bagave2022accountable}, highlighting the multifaceted nature of accountability.

\subsubsection{Accountability and Transparency}

The relationship between accountability and transparency is symbiotic. Transparency is not an adjunct but a foundational element that empowers accountability.  Effective accountability is contingent upon the disclosure of relevant information, enabling stakeholders to scrutinize, understand, and, if necessary, contest decisions. This necessitates the provision of comprehensive provenance documentation, elucidating the design, development and decision-making pathways \cite{konigstorfer2022ai}.
Moreover, transparency mandates the explicit delineation of responsibilities, which is vital to prevent the dilution of accountability and to ensure that responsible parties can be identified and held to account, especially from a cross-organizational supply chain perspective  \cite{cobbe2023understanding} where there could be shared accountability.


\subsection{AI Accountability}
Accountability within AI systems is not merely an extension of traditional accountability concepts but a distinct challenge due to the autonomous and often opaque nature of AI decision-making processes. As AI becomes more integrated into societal functions, the imperative for robust accountability frameworks is underscored by a range of RAI guidelines and principles \cite{jobin2019global, ratzan2023measuring}.
Examples include both organizational (e.g., Microsoft \cite{MicrosoftRAIPrinciple}, OECD \cite{OECDRAIPrinciple}) and governmental (e.g., Australia \cite{AU_Principle}, US \cite{NIST_AIRMF}, EU \cite{EUList2020}) efforts.


\subsubsection{AI Accountability Challenge}

Built on Nissenbaum \cite{nissenbaum1996accountability}, Cooper et al. \cite{cooper2022accountability} revisited the barriers to algorithmic accountability:

\textbf{Many Hands}: The involvement of multiple parties in AI development and use, including reliance on third-party tools, disperses responsibility and complicates accountability.

\textbf{``Bugs''}: AI ``bugs'' extend beyond typical software errors, encompassing data biases, modeling errors, and design flaws, often systemic and unpredictable due to ML's statistical nature.

\textbf{The ``computer'' as scapegoat}: Misattributing moral agency to AI systems obscures human responsibility, complicating meaningful accountability.

\textbf{Ownership without liability}: Firms often claim rights over AI assets while avoiding responsibility for their impacts, impeding transparency and independent auditing. This calls for stronger legal and ethical frameworks to ensure liability, such as \cite{EU_liability}.

\subsubsection{Accountability Challenge in the GenAI Era}

GenAI models, characterized by their extensive scale, complexity, and adaptability, introduces new dimensions to these established challenges:

\textbf{``Many \textit{More} Hands''}: The collaborative nature of GenAI model development, involving a broad spectrum of contributors from individuals to multinational corporations, significantly disperses responsibility. This dispersion is intensified when third parties modify or tailor these models for specific applications. The distribution of GenAI models via APIs or cloud platforms adds another layer of complexity to the accountability equation.

\textbf{``Bugs'' at Scale}: The scale and complexity of GenAI models exacerbate the consequences of such ``bugs'', transforming them from mere technical faults into systemic challenges. This means that biases, errors, and unpredictable behaviors can have far-reaching consequences. 
These issues complicate AI accountability, as the probabilistic nature of GenAI models often precludes clear-cut explanations for their outputs.

\textbf{The ``Big Black Box'' as Scapegoat}: The complex nature of large GenAI models often leads to anthropomorphization, where these systems are perceived as more human-like. This perception, combined with their ``black box'' nature, can be misleadingly used to deflect accountability from human actors to the AI systems. 

\textbf{Ownership with Disavowed Liability}: The proprietary nature of many GenAI models often leads to a dichotomy where organizations claim credit for successes but not liability for failures. This issue is compounded by the wide application of GenAI models and the lack of transparency that hampers independent auditing efforts.


\subsection{Related Work}
\textbf{RAI Operationalization}.
Operationalizing RAI has garnered attention in both industry and academia, with a notable shift towards process-centric approaches. Singapore's AI Verify \cite{SA_AIVerify} stands out for its evaluation of AI systems against 12 RAI principles, including accountability, through a blend of process checks and technical tests. The EU's capAI project \cite{floridi2022capai} is pioneering a conformity assessment procedure aligned with the EU AI Act. Credo AI \cite{CredoAI2023} integrates process checks into its AI governance platform, aligning with policy requirements. UC Berkeley's taxonomy \cite{BerkeleyCLTC2023AITrustworthiness} aligns with US National Institute of Standards and Technology (NIST) AI RMF's RAI principles, focusing on organizational processes. Fraunhofer IAIS's work \cite{BSI2021AIC4} provides concrete criteria for RAI assessment, while Raja et al. \cite{ratzan2023measuring} propose processes for RAI assessment in banking.
More specifically in AI accountability, however, discourse predominantly remains conceptual, with studies (e.g., \cite{doshi2017accountability, raja2023ai, smith2021clinical, busuioc2021accountable}) focusing on high-level theoretical aspects.

\textbf{AI Measurement and Metrics}
The field of RAI measurement and metrics is rapidly evolving. Research on AI explainability \cite{islam2022systematic} and fairness \cite{castelnovo2022clarification,franklin2022ontology} typically targets model-level requirements, highlighting a gap in system-level analysis. Notably, NIST's work \cite{NIST_AIMeasurement} on AI metrics emphasizes model-level metrics like accuracy and bias, ignoring accountability. The OECD's RAI Metrics catalogue \cite{OECD2023TrustworthyAIMetrics}, while comprehensive, remains focused on model-level technical requirements and lacks metrics for accountability. The catalogue has specific metrics related to content generation tasks. Stanford's Foundation Model Transparency Index \cite{bommasani2023foundation} offers a broader range of system-level metrics for assessing foundation model transparency, spanning from model construction to downstream usage.

Our research diverges by focusing on concrete, process-centric metrics for AI accountability, complemented by resource and product metrics (see Table \ref{tab:comparison}). This approach offers system-level metrics applicable to both traditional AI and GenAI systems, providing concrete guidance for operationalizing AI accountability.

\begin{table}[]
\centering
\caption{Comparison of Metrics Related Work}
\label{tab:comparison}
\resizebox{0.8\columnwidth}{!}{%
\begin{tabular}{|l|l|l|l|l|l|}
\hline
\textbf{} & \textbf{NIST \cite{NIST_AIMeasurement}} & \textbf{OECD \cite{OECD2023TrustworthyAIMetrics}} & \textbf{Stanford \cite{bommasani2023foundation}} & \textbf{AI Verify \cite{SA_AIVerify}} & \textbf{This Work} \\ \hline
\textbf{Focus} & Model-level & Model-level & \textbf{System-level} & \textbf{System-level} & \textbf{System-level} \\ \hline
\textbf{Accountability} & No & No & No & \textbf{Yes} & \textbf{Yes} \\ \hline
\textbf{GenAI} & No & Partly & \textbf{Yes}, GenAI only & No & \textbf{Yes} \\ \hline
\end{tabular}
}
\end{table}


\section{Methodology}
\begin{figure}[ht]
  \centering
  \includegraphics[width=0.6\linewidth]{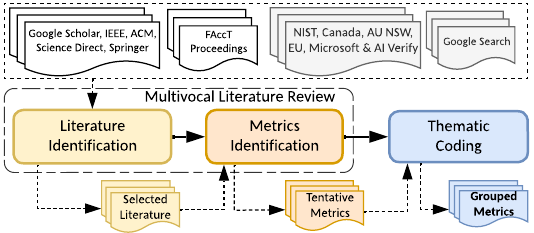}
  \caption{Methodology Overview}
  \label{Fig:method}
\end{figure}

Our research methodology unfolds in two sequential phases, as depicted in Fig. \ref{Fig:method}: a Multivocal Literature Review (MLR) followed by thematic coding. The MLR, adhering to established guidelines \cite{Garousi_2019, kitchenham2009systematic}, merges Systematic Literature Review (SLR) and Grey Literature Review (GLR) to ensure a holistic understanding of AI accountability practices. This blended approach mitigates the limitations inherent in using either SLR or GLR in isolation. While SLR alone may overlook emerging trends and practical insights, GLR exclusively might lack the methodological rigor of peer-reviewed academic literature. Our methodology, therefore, balances these aspects to provide a comprehensive and robust analysis.

\subsection{Multivocal Literature Review (MLR)}

\subsubsection{Systematic Literature Review (SLR)}

\textbf{Search Strategy}: Our keyword choices aimed to ensure comprehensive coverage of the domain. The search string used was: \textit{(AI OR ML OR ``artificial intelligence'' OR ``machine learning'' OR ``large language model'' OR LLM OR ``Foundation Model'' OR ``Frontier Model'' OR ``Generative AI'') AND (accountability OR accountable)}. Pilot tests ensured their suitability. The search string was used to search title only. Searches were conducted on 26 September 2023.
Searches spanned five databases: IEEE Xplore, ACM Digital Library, Science Direct, Springer, and Google Scholar, yielding 221 results.
Additionally, we reviewed ACM Conference on Fairness, Accountability, and Transparency proceedings given its relevance to our research focus, focusing on titles including "Accountability" or "Accountable," resulting in 33 papers. The search was not constrained by publication date.

\textbf{Inclusion \& Exclusion, and Quality Assessment}: We included sources based on: 1) their relevance to AI accountability after initial Abstract screening; 2) the inclusion of accountability practices/processes based on in-depth review. Exclusions were made for non-English sources and conference papers superseded by journal versions. he quality assessment was thorough, emphasizing the clarity of research aims, robustness of design, and the articulation of findings, contributions, and limitations. This rigorous process resulted in the selection of 40 pertinent studies. The distribution of these studies over recent years is as follows: 6 published in 2023, 14 in 2022, 12 in 2021, and 8 between 2017 and 2020, underscoring the growing interest and attention in AI accountability.

\textbf{Data Extraction}: Data extraction was conducted manually, utilizing Zotero for review and annotation. We extracted information on processes/practices for AI accountability (process metrics), tools and resources for AI accountability (resource metrics), and the resultant artefacts (product metrics). The extraction of resource and product metrics was contingent upon the identification of relevant process metrics.

\subsubsection{Grey Literature Review (GLR)}

\textbf{Search Strategy}: Employing the same keywords as in SLR, our GLR involved searches on Google Search Engine, yielding 146 results. We also reviewed six internationally recognized AI governance/risk management frameworks based on their quality and representativeness \cite{lee2023qb4aira}: EU Trustworthy AI Assessment List \cite{EUList2020}, NIST AI RMF \cite{NIST_AIRMF}, Canada Algorithmic Impact Assessment Tool \cite{CAAIA}, Australia NSW AI Assurance Framework \cite{NSW}, Microsoft Responsible AI Impact Assessment Template \cite{MS_RAItemplate}, and Singapore's AI Verify \cite{SA_AIVerify}.

\textbf{Inclusion and Exclusion Criteria \& Quality Assessment \& Data Extraction}: The GLR adhered to the same rigorous criteria of Inclusion/Exclusion and Quality Assessment as the SLR, with an added emphasis on the credibility of grey literature sources \cite{Garousi_2019}. This led to an initial selection of 7 studies, which was expanded to 24 through snowballing, acknowledging the interconnected nature of grey literature. The extracted data also focused on the same three types of metrics as in SLR. Similar to SLR, 5 of the studies were published in 2023, 7 in 2022, 4 in 2021, and 7 between 2018-2020, with 1 exception with no clearly stated publication date.

\subsection{Thematic Coding}

Our thematic coding process employed a hybrid approach, blending both deductive and inductive methods \cite{fereday2006demonstrating}. This approach leverages the strengths of both structured and emergent analysis, ensuring a comprehensive understanding of the data. This hybrid thematic coding approach allowed us to systematically categorize the extracted metrics into a structured yet flexible framework. The predefined criteria provided a clear direction for our analysis, while the emergent sub-criteria offered depth and detailed insights, leading to a robust and comprehensive understanding of AI accountability.

\begin{enumerate}
\item \textbf{Deductive Coding}: We began with three predefined broad criteria (i.e., Responsibility, Auditability, Redressability), as described in Section \ref{sec:AccConcept}. These criteria served as initial guides for categorizing the extracted metrics.
\item \textbf{Emergent Sub-Criteria}: As we delved deeper into the coding process, we allowed for the emergence of sub-themes. These sub-themes were not predefined but were identified based on patterns, similarities, and differences in the data. This inductive aspect of our analysis allowed for a nuanced understanding of the metrics and their interrelations.

\item \textbf{Refinement and Integration}: The emergent sub-criteria were continuously refined and integrated into the broader predefined themes. This iterative process ensured that our thematic structure accurately represented the complexities and nuances of the data.

\item \textbf{Internal Validation}: To ensure the validity of our thematic structure, one author conducted the coding and three authors reviewed the results and provided feedback. Adjustments were made to make sure consensus among authors.

\end{enumerate}

Upon completion of the MLR and thematic coding, our analysis distilled eleven key process metrics systematically categorized into five emergent sub-criteria, which align with three pre-defined overarching pillar criteria, as depicted in Table \ref{tab:catalog}.

\begin{table*}[]
\caption{System-Level Metrics Catalogue for AI Accountability}
\label{tab:catalog}
\resizebox{\textwidth}{!}{%
\begin{tabular}{|c|l|l|l|l|l|}
\hline
\multicolumn{1}{|l|}{\textbf{Criteria}} & \textbf{Sub-Criteria}                 & \textbf{Process Metrics}         & \textbf{Key Considerations}                                                                                                                                                                                                                                                                                   & \textbf{Resource Metrics}                                                                                                                                                                                                                                                                               & \textbf{Product Metrics}                                                                                                                                                        \\ \hline
\multirow{5}{*}[-1.5cm]{Responsibility}         & \multirow{3}{*}[-0.5cm]{RAI Oversight}        & Roles and Responsibilities       & \begin{tabular}[c]{@{}l@{}}\textbullet\hspace{2pt}Comprehensive role clarity: \\  \hspace{6pt}-\hspace{2pt}Design and development\\  \hspace{6pt}-\hspace{2pt}Deployment and operations\\  \hspace{6pt}-\hspace{2pt}Procurement and integration\\  \hspace{6pt}-\hspace{2pt}Governance and compliance\\  \hspace{6pt}-\hspace{2pt}AI as a service\end{tabular}                                                                                         & \multirow{5}{*}[-1cm]{\begin{tabular}[c]{@{}l@{}}\textbullet\hspace{2pt}Soft laws (e.g., best\\ \hspace{5pt}practices, guidelines\\ \hspace{5pt}standards etc)\\  \textbullet\hspace{2pt}Hard laws (e.g., EU\\ \hspace{5pt}AI Act)\end{tabular}}                                                                                                                                   & \begin{tabular}[c]{@{}l@{}}\textbullet\hspace{2pt}Procedure Manuals\\ \textbullet\hspace{2pt}Contracts or agreements\\ \textbullet\hspace{2pt}Position descriptions\\ \textbullet\hspace{2pt}Recruitment practices\\ \textbullet\hspace{2pt}Workforce dev strategy\end{tabular} \\ \cline{3-4} \cline{6-6} 
                                        &                                       & RAI Governance Committee         & \begin{tabular}[c]{@{}l@{}}\textbullet\hspace{2pt}Multidisciplinary composition\\ \textbullet\hspace{2pt}Strategic leadership involvement\end{tabular}                                                                                                                                                                                                  &                                                                                                                                                                                                                                                                                                & \textbullet\hspace{2pt}Policy doc on Committee                                                                                                                                              \\ \cline{3-4} \cline{6-6} 
                                        &                                       & Organizational AI Risk Tolerance & \begin{tabular}[c]{@{}l@{}}\textbullet\hspace{2pt}Tiered risk-based categorization\\ \textbullet\hspace{2pt}Balancing competing interests\end{tabular}                                                                                                                                                                                                        &                                                                                                                                                                                                                                                                                                & \begin{tabular}[c]{@{}l@{}}\textbullet\hspace{2pt}Policy doc on org's risk\\ \hspace{5pt}tolerance and mitigations\end{tabular}                                                                         \\ \cline{2-4} \cline{6-6} 
                                        & \multirow{2}{*}[-0.5cm]{RAI Competence}       & RAI Training                     & \begin{tabular}[c]{@{}l@{}}\textbullet\hspace{2pt}Holistic training content\\ \textbullet\hspace{2pt}Targeted training for diverse roles\\ \textbullet\hspace{2pt}Adaptive and ongoing education\end{tabular}                                                                                                                                                                &                                                                                                                                                                                                                                                                                                & \textbullet\hspace{2pt}Training certificates                                                                                                                                                \\ \cline{3-4} \cline{6-6} 
                                        &                                       & RAI Capability Assessment        & \begin{tabular}[c]{@{}l@{}}\textbullet\hspace{2pt}Multifaceted assessment\\ \textbullet\hspace{2pt}Standard alignment\\ \textbullet\hspace{2pt}Organizational RAI maturity\\ \textbullet\hspace{2pt}Continuous enhancement\end{tabular}                                                                                                                                                           &                                                                                                                                                                                                                                                                                                & \textbullet\hspace{2pt}Assessment reports                                                                                                                                                   \\ \hline
\multirow{4}{*}[-2.5cm]{Auditability}           & \multirow{3}{*}[-2cm]{Systematic Oversight} & Data Provenance                  & \begin{tabular}[c]{@{}l@{}}\textbullet\hspace{2pt}Detailed data record-keeping\\ \textbullet\hspace{2pt}Data version control\\ \textbullet\hspace{2pt}Data integrity and risk mitigation\\ \textbullet\hspace{2pt}Legal and ethical compliance\end{tabular}                                                                                                                                       & \multirow{4}{*}[-1cm]{\begin{tabular}[c]{@{}l@{}}\textbullet\hspace{2pt}Soft laws (e.g.,\\ \hspace{5pt}auditing guidelines\\ \hspace{5pt}and frameworks etc)\\  \textbullet\hspace{2pt}Hard laws (e.g., EU\\ \hspace{5pt}AI Act)\\  \textbullet\hspace{2pt}AI documentation\\ \hspace{5pt}tools (e.g., datasheets,\\ \hspace{5pt}model/system cards)\\ \textbullet\hspace{2pt}Technical tools (e.g., \\ \hspace{5pt}blockchain, \\ \hspace{5pt}knowledge graph)\end{tabular}} & \multirow{2}{*}{\begin{tabular}[c]{@{}l@{}}\textbullet\hspace{2pt}Provenance records\\ \textbullet\hspace{2pt}System features (e.g., auto-\\ \hspace{5pt}logging, version control\end{tabular}}                              \\ \cline{3-4}
                                        &                                       & Model Provenance                 & \begin{tabular}[c]{@{}l@{}}\textbullet\hspace{2pt}Detailed model record-keeping\\ \textbullet\hspace{2pt}Model selection and validation\\ \textbullet\hspace{2pt}Model version control\end{tabular}                                                                                                                                                                          &                                                                                                                                                                                                                                                                                                &                                                                                                                                                                        \\ \cline{3-4} \cline{6-6} 
                                        &                                       & System Provenance and Logging    & \begin{tabular}[c]{@{}l@{}}\textbullet\hspace{2pt}Detailed system record-keeping\\     \hspace{6pt}-\hspace{2pt}System version control\\     \hspace{6pt}-\hspace{2pt}Decision/Trade-off\\ \textbullet\hspace{2pt}Comprehensive operational logging\\   \hspace{6pt}-\hspace{2pt}User interaction and system response\\   \hspace{6pt}-\hspace{2pt}Incident and response\\   \hspace{6pt}-\hspace{2pt}System configuration changes\\ \textbullet\hspace{2pt}Composition Management\end{tabular} &                                                                                                                                                                                                                                                                                                & \begin{tabular}[c]{@{}l@{}}\textbullet\hspace{2pt}Provenance records\\ \hspace{5pt}(and logs)\\ \textbullet\hspace{2pt}System features (e.g., auto-\\ \hspace{5pt}logging, version control\end{tabular}                                  \\ \cline{2-4} \cline{6-6} 
                                        & Compliance Checking                   & Auditing                         & \begin{tabular}[c]{@{}l@{}}\textbullet\hspace{2pt}Diversified auditing strategy\\  \textbullet\hspace{2pt}Multi-dimensional audit techniques\\  \textbullet\hspace{2pt}Ethical and legal compliance\\  \textbullet\hspace{2pt}Regular audits \\  \textbullet\hspace{2pt}Verifiable audits\\  \textbullet\hspace{2pt}Audit-driven improvements\end{tabular}                                                                                  &                                                                                                                                                                                                                                                                                                & \begin{tabular}[c]{@{}l@{}}\textbullet\hspace{2pt}Audit reports\\ \textbullet\hspace{2pt}Compliance certificates\\ \hspace{5pt}and licenses\end{tabular}                                                                      \\ \hline
\multirow{2}{*}[-0.2cm]{Redressability}         & \multirow{2}{*}[-0.2cm]{Redress-by-Design}    & Incident Reporting and Response  & \begin{tabular}[c]{@{}l@{}}\textbullet\hspace{2pt}Accessibility and Visibility\\ \textbullet\hspace{2pt}Structured Incident Management\\ \textbullet\hspace{2pt}Feedback Loop Integration\end{tabular}                                                                                                                                                                       & \multirow{2}{*}[0.2cm]{\begin{tabular}[c]{@{}l@{}}\textbullet\hspace{2pt}Redundancy design\\ \hspace{5pt}case studies\\ \textbullet\hspace{2pt}Incident \\ \hspace{5pt}management tools\end{tabular}}                                                                                                                                                                    & \begin{tabular}[c]{@{}l@{}}\textbullet\hspace{2pt}Incident and response doc\\ \textbullet\hspace{2pt}System features (user\\ \hspace{5pt}feedback and report)\end{tabular}                                                   \\ \cline{3-4} \cline{6-6} 
                                        &                                       & Built-in Redundancy              & \textbullet\hspace{2pt}Multi-Modal Redundancy                                                                                                                                                                                                                                                                                      &                                                                                                                                                                                                                                                                                                & \begin{tabular}[c]{@{}l@{}}\textbullet\hspace{2pt}System features (redundant\\ \hspace{5pt}components/functionalities)\end{tabular}                                                                     \\ \hline
\end{tabular}%
}
\end{table*}

\section{Results - Responsibility}

\label{Sec:Responsibility}
Responsibility serves as a cornerstone of accountability in AI governance, specifying who is answerable for each phase of an AI system's lifecycle. It goes beyond mere task allocation to mandate that these responsible entities possess the necessary ethical awareness, technical knowledge, and expertise. This ensures competent decision-making, thereby strengthening accountability.

\subsection{Sub-Criterion 1.1: RAI Oversight}
This emphasizes the necessity for a well-defined \textit{organizational structure} to oversee the AI lifecycle, from procurement and development to deployment and operations. Such a structure is pivotal in upholding high ethical standards and responsibility in AI solutions, both in-house and externally sourced.

\subsubsection{\faCog\,\textbf{Process Metric 1.1.1: Roles and Responsibilities}}
\textbf{Context and Importance}:
The clear delineation of roles and responsibilities is critical for the ethical, responsible, and effective development, deployment, procurement, and governance of AI systems \cite{OECD_AdvancingAIAcc, smith2021clinical, matthews2020patterns}. This clarity is not only vital for defining responsibilities but also for managing potential overlaps and conflicts among roles. By proactively addressing these challenges, organizations can ensure cohesive and effective operation across all stages of the AI lifecycle. This approach is universally necessary, extending from traditional AI to GenAI. 

\textbf{Key Consideration-Comprehensive Role Clarity}:

\begin{itemize}
    \item \textbf{Design and Development}: Formalize roles associated with the initial stages of AI systems, covering plan and design, data collection and (pre)processing, model development, comprehensive testing (validation and verification), and fine-tuning in the case of large-scale GenAI models. 
    
    \item \textbf{Deployment and Maintenance}: Define roles pertinent to the deployment, system monitoring, and operational management (e.g., issue resolution, upgrades) \cite{MsRAIStandard}.
    
    \item \textbf{Procurement and Integration}: Establish clear roles for evaluating, selecting, and integrating AI systems, focusing on legal compliance, security, and ethical alignment with organizational values \cite{DOEPlaybook, MsRAIStandard}. These should include both technical (e.g., developers) and non-technical (e.g., procurement specialists) roles.
    
    \item \textbf{Governance and Compliance} \cite{fetic2020principles, DOEPlaybook}: Assign roles for policy development, ethical standards adherence, and regulatory compliance. Include roles for internal and external audits to ensure continuous ethical assessment.
    
    \item \textbf{Service Provision (AI as a Service)}:
    For AI systems offered as a service  \cite{javadi2020monitoring}, formalize roles for managing shared responsibilities and accountability \cite{cobbe2023understanding}, particularly for GenAI, ensuring clear agreements on usage policies \cite{OpenAIComment23}, data governance, and liability.
\end{itemize}

\subsubsection{\faCog\,\textbf{Process Metric 1.1.2: AI Governance Committee}}
\textbf{Context and Importance}:
The establishment of an AI Governance Committee, or a similar oversight body, is ethically and effectively managing AI systems ~\cite{memarian2023fairness}, especially in the dynamic and complex realm of GenAI \cite{fui2023generative}. This committee should oversee AI systems throughout their lifecycle, from conception to decommissioning, ensuring consistent and responsible governance \cite{HKPCPD2021}.
Its role extends to efficient decision-making, particularly in ethical and risk management scenarios, necessitating regular collaboration with internal and external stakeholders, including regulatory bodies, to stay aligned with best practices and evolving regulations.

\textbf{Key Considerations}:

\begin{itemize}
\item \textbf{Multidisciplinary Composition}: The committee should comprise experts from diverse fields like law, ethics, academia, and technology \cite{Reid_AICommittee}. This diversity enables a comprehensive approach to AI risk management, facilitating the reconciliation of varied perspectives and addressing the complex ethical and operational challenges of (Gen)AI systems \cite{DOEPlaybook}.

\item \textbf{Strategic Leadership Involvement}: Incorporating senior leadership in the committee, akin to models like Microsoft’s Responsible AI Council \cite{MSComment23}, is crucial. This ensures that strategic decisions are informed by a wide range of organizational perspectives and are aligned with RAI practices.
\end{itemize}






\subsubsection{\faCog\,\textbf{Process Metric 1.1.3: Organizational AI Risk Tolerance}}
\textbf{Context and Importance}:
Defining an organizational risk tolerance/appetite is a critical aspect of risk-based approach to AI risk management \cite{NIST_AIRMF}, particularly for GenAI. This risk tolerance delineates the operational boundaries for AI systems, aligning them with ethical standards and operational requirements. It is essential for navigating ethical dilemmas like balancing fairness with accuracy and managing data privacy across different regulatory landscapes. The EU AI Act's \cite{EUAIACT} risk classification framework (i.e., Unacceptable/High/Limited/Minimal or no risk) offers a valuable model for categorizing AI systems based on risk levels and imposing transparency obligations specific to GenAI.

\textbf{Key Considerations}:

\begin{itemize}
    \item \textbf{Tiered Risk-based Categorization}: Adopting a tiered risk-based categorization, aligned with standards and regulations like the EU AI Act, is recommended. This categorization should consider the nature, scope, and purpose of the AI system, as well as the organization's unique circumstances. For instance, healthcare AI systems may have different risk tolerances than retail AI, reflecting their varying ethical and privacy concerns.
    \item \textbf{Balancing Competing Interests}: It's crucial to balance competing interests such as data protection and operational efficiency.While data protection is paramount, overly rigid policies can stifle innovation. Organizations should aim for a balance that accommodates both ethical considerations and operational efficiency \cite{UKICO20}.
\end{itemize}




\subsection{Sub-Criterion 1.2: RAI Competence}
RAI Competence is essential for individuals and organizations to effectively implement and uphold RAI principles. In the GenAI context, this need is amplified due to its unique challenges and the fast pace of technological evolution. Ensuring RAI understanding across all organizational levels, from technical staff to executive leadership, is critical for maintaining RAI practices.

\subsubsection{\faCog\,\textbf{Process Metric 1.2.1: RAI Training}}
\textbf{Context and Importance}: 
Effective RAI training is indispensable for fostering an organizational awareness and culture that prioritizes RAI development and deployment \cite{OECD_AdvancingAIAcc, raja2023ai, EUList2020, bogina2021educating}. 
In GenAI, the high stakes associated with potential missteps due to advanced capabilities necessitate comprehensive and continuous training. This training should address the current state of technology and ethics while anticipating future developments and challenges, ensuring that all stakeholders are prepared to make responsible decisions in a rapidly changing environment.

\textbf{Key Considerations}:

\begin{itemize}
\item \textbf{Holistic Training Content}: Training should be holistic \cite{raja2023ai}, including technical aspects, ethical considerations, legal compliance, and risk management. Specific modules could include data privacy, algorithmic fairness, and legal compliance.

\item \textbf{Targeted Training for Diverse Roles}: Training should be customized to suit different roles within the organization. For instance, legal professionals may focus on compliance, while developers may delve into ethical coding practices \cite{HKPCPD2021}. Additional training should also be extended to procurement specialists and executive leadership, focusing on governance, risk management, and control \cite{DOEPlaybook}. This approach equips various stakeholders with the skills necessary to cultivate organizational RAI competence.

\item \textbf{Adaptive and Ongoing Education} \cite{GoogleComment23, solyst2023would}: The field of AI is continuously evolving, making ongoing education and adaptation essential. Training programs should not only include regular updates and feedback mechanisms but also adapt to new ethical challenges, technological advancements, and diverse global perspectives (i.e., future-proof).
\end{itemize}



\subsubsection{\faCog\,\textbf{Process Metric 1.2.2: RAI Capability Assessment}}
\textbf{Context and Importance}: Transitioning from being merely ``trained'' in RAI to being ``competent'' in their implementation is a critical evolution for organizations. This transition requires robust mechanisms to evaluate and certify individual staff competencies in RAI, as well as to assess and enhance the overall organizational RAI maturity. Such an approach ensures that both individuals and the organization are proficient in implementing RAI effectively, with a focus on empowering staff with necessary skills and equipping the organization for RAI practices.

\textbf{Key Considerations}:

\begin{itemize}
\item \textbf{Multifaceted Assessment}: Implement a variety of assessment methods to appraise RAI competence across diverse roles, including exams, practical exercises, and scenario analyses, covering both technical and ethical aspects of AI \cite{raja2023ai}. 
For instance, developers may be assessed on technical RAI knowledge, while legal professionals may be evaluated on compliance related aspects. This ensures a robust, role-specific measure of RAI competence.

\item \textbf{Standard Alignment and Benchmarking}: The criteria and methodologies for assessment should align with evolving industry standards and best practices in AI ethics, drawing on guidelines from bodies like the IEEE SA \cite{IEEE2023AIStandards}.

\item \textbf{Organizational Readiness and RAI Maturity}:
Adopting or adapting an RAI maturity model (e.g., Microsoft RAI Maturity Model \cite{Vorvoreanu2023RAIMaturityModel}) is critical to gauge and enhance the organization’s overall capability in implementing RAI \cite{lupattern}. This model should evaluate aspects such as governance, technical infrastructure, ethical alignment, and stakeholder engagement, serving as a tool for both diagnosis and continuous improvement.

\item \textbf{Continuous Policy and Training Enhancement}: Leverage insights from the assessments to continuously refine training programs and organizational policies, ensuring they remain relevant in the rapidly changing AI landscape.
\end{itemize}



\subsection{Resource \& Product Metrics -- Responsibility}

\textit{Responsibility} necessitates a strategic approach at the organizational level, integrating both resources and products to ensure ethical AI development and management. Resources provide the necessary tools and frameworks for ethical AI governance, while products represent the tangible outcomes of these efforts.

\faLightbulbO\,Essential \textbf{\textit{Resources}} for \textit{Responsibility}: 

\begin{itemize}

    \item \textbf{Soft Laws}: These encompass AI ethics best practices, guidelines, governance frameworks, and standards of care \cite{cooper2022accountability, MSComment23}, including both general principles and specific directives, such as procurement guidelines and aforementioned RAI maturity models. Examples include US NIST AI RMF \cite{NIST_AIRMF} and IEEE 7000 Standard Series \cite{IEEE2023AIStandards}.

    \item \textbf{Hard Laws}: Adherence to legal frameworks is essential, especially for regulated sectors. These laws establish a legal baseline for responsible AI practices, ensuring compliance and ethical integrity.Legislation and regulations are crucial, especially for organizations in regulated sectors. Examples include the EU AI Act \cite{EUAIACT} and the US Blueprints for an AI Bill of Rights \cite{WhiteHouse2023AIBillOfRights}.
\end{itemize}

\faPuzzlePiece\,\textbf{\textit{Products}}: The implementation of \textit{Responsibility} requires the development of specific organizational policy \textbf{documentation} \cite{SA_AIVerify} related to different processes:

\begin{itemize}
    \item \textbf{Roles and Responsibilities}: This includes internal procedure manuals, contracts, and written agreements that clearly delineate roles and responsibilities in AI system development and deployment \cite{SA_AIVerify, EUList2020, novelli2023accountability, markovic2021accountability, fetic2020principles}. It also encompasses position descriptions, recruitment practices, and workforce development strategies \cite{DOEPlaybook}. These documents are essential for establishing clear accountability and ethical guidelines within the organization.
    \item \textbf{AI Governance Committee}: Detailed documentation outlining the structure, function, and protocols of the Committee \cite{SA_AIVerify}. This includes its role in decision-making, oversight, and ensuring compliance with ethical standards.
    \item \textbf{Organizational AI Risk Tolerance}: Documents defining the organization's AI risk tolerance levels and mitigation strategies are crucial. This encompasses the assessment and management of different risk scenarios, aligning AI initiatives with the organization's overall risk framework.
    \item \textbf{RAI Training}: Certificates or other formal documents that validate the successful completion of RAI training, serving as a testament to the individual's proficiency in RAI.
    \item \textbf{RAI Capability Assessment}: Comprehensive reports that evaluate the RAI competencies at both individual and organizational levels, highlighting areas of strength and opportunities for improvement.
\end{itemize}
Weaving all these for \textit{RAI Oversight} together, the organization can establish its own \textbf{RAI framework and structures}.

\section{Results - Auditability}
\label{Sec:Auditability}
\textit{Auditability} is an indispensable aspect that bolsters accountability in AI. It enables thorough inspection and evaluation of AI systems, ensuring they conform to established standards and objectives. Auditability facilitates transparency and enables rigorous assessment, substantiating compliance and enhancing the ethical integrity of AI systems.

\subsection{Sub-Criterion 2.1: Systematic Oversight}

\textit{Systematic Oversight} represents a comprehensive record-keeping and logging framework. It encompasses not just data and model lineage but extends to the entire AI system, encompassing development processes and operational dynamics. This holistic oversight is instrumental in maintaining system integrity and providing stakeholders with essential information for in-depth audits, ensuring every aspect of AI development and operation is subject to scrutiny.


\subsubsection{\faCog\,\textbf{Process Metric 2.1.1: Data Provenance}}

\textbf{Context and Importance}:
Data Provenance is critical in AI risk management and governance, particularly for GenAI, where it bolsters transparency and ethical compliance. The dependency of GenAI on extensive datasets not only amplifies its importance but also presents unique challenges. Issues such as copyright, fair use, and privacy risks, as identified by Khan and Hanna \cite{khan2022subjects}, accentuate the necessity for robust Data Provenance. It's essential in verifying data integrity and ethical utilization in (Gen)AI systems, addressing these challenges to uphold the trustworthiness of AI.

\textbf{Key Considerations}:

\begin{itemize}

\item \textbf{Detailed Data Record-Keeping} \cite{naja2022using, matthews2020patterns, peregrina2022towards, MSComment23, gursoy2022system}: Implement a thorough record-keeping system that documents all aspects of the data lifecycle, including sources, dataset characteristics, collection methods, preprocessing steps, and usage patterns. This comprehensive approach ensures that every aspect of data handling is auditable.

\item \textbf{Data Version Control} \cite{SA_AIVerify}: Implement a robust data version control system to track changes and updates in the data used over time. This system should document each version's specific characteristics and modifications, providing clarity on the evolution of the data set.

\item \textbf{Data Integrity and Risk Management} \cite{floridi2022capai, khan2022subjects, naja2022using}: Ensure that the data used in AI systems is scrutinized for quality factors like representativeness, relevance, and accuracy and proactively address data-related risks.

\item \textbf{Ethical and Legal Compliance}  \cite{khan2022subjects, floridi2022capai}: The data provenance process could incorporate thorough assessments and documentation to ensure that an organization's data practices adhere to established ethical expectations and legal standards, like the EU GDPR.

\end{itemize}



\subsubsection{\faCog\,\textbf{Process Metric 2.1.2: Model Provenance}}

\textbf{Context and Importance}:
Model Provenance, especially for GenAI, is indispensable for maintaining ethical integrity and transparency. The inherent complexity of GenAI models, characterized by sophisticated algorithms, necessitates meticulous documentation of their development and deployment processes. This is crucial for ensuring that (Gen)AI models adhere to ethical standards and regulatory norms throughout their lifecycle.

\textbf{Key considerations}:

\begin{itemize}

\item \textbf{Detailed Model Record-Keeping} \cite{naja2022using, matthews2020patterns, zainyte2021challenges, GoogleComment23, floridi2022capai}: Maintain extensive records that capture model characteristics, algorithmic details, decision-making thresholds, and data pathways leading to final decisions is essential. This documentation should include intended and unintended usages, known limitations, and associated risks.

\item \textbf{Model Selection and Validation} \cite{raja2023ai, GoogleComment23}: Document the rationale for model selection, with an emphasis on fairness, explainability, and robustness, which are particularly pertinent in GenAI. A well-defined validation strategy is also essential, ensuring clear accountability for model design and implementation.

\item \textbf{Model Version Control} \cite{SA_AIVerify}: Implement systematic version control is critical for tracking changes and updates made to the model over time, including each version’s specific features and modifications.

\end{itemize}



\subsubsection{\faCog\,\textbf{Process Metric 2.1.3: System Provenance and Logging}}

\textbf{Context and Importance}:
System Provenance and Logging embodies a holistic approach to understanding and overseeing AI systems at a systemic level, including both AI and non-AI elements \cite{markovic2021accountability, gursoy2022system}. This consolidates the comprehensive documentation of an AI system's developmental history with the nuanced monitoring of its operational dynamics. It ensures a profound grasp of the entire lifecycle of AI systems, from their inception and architectural evolution to their real-time responses and operational adjustments.

\textbf{Key Considerations}:

\begin{itemize}

\item \textbf{Detailed System Record-Keeping} \cite{zainyte2021challenges, OpenAIComment23, akhgar2022accountability, cooper2021accuracy, fetic2020principles}: 
    \begin{itemize}
        \item \textbf{System Version Control}: Thorough records of the system's architecture and components, documenting the evolution of both AI and non-AI elements. Document their evolution (e.g., code versioning) over time.
        \item \textbf{Decision/Trade-off}: Capture decisions made by human agents and automated processes that have influenced the overall AI system. Clarify the causal relationships between different stages of system development to provide a clear understanding of how decisions impact system behavior and outcomes.
    \end{itemize}

\item \textbf{Comprehensive Operational Logging} \cite{henriksen2021situated, Microsoft2023LogMonitorAzureOpenAI, DOEPlaybook, CAAIA}:
    \begin{itemize}
        \item \textbf{User Interactions and System Responses}: Log all user queries/prompts and system responses to track usage patterns and system performance.
        \item \textbf{Incident and Response Logging}: Maintain a detailed log of operational incidents, including system errors or failures, and document the steps taken for resolution. This log should align with the organization's risk management and incident response strategies. 
        \item \textbf{System Configuration Changes}: Keep records of all changes made to the system's configuration and operational parameters in real-time, ensuring a clear trail for audit and review.
\end{itemize}
\item \textbf{Composition Analysis and Vulnerability Management} \cite{DOEPlaybook}: 
Establish mechanisms for the systematic identification and analysis of AI components, complementing traditional software composition analysis. This process should attend to the unique challenges inherent in integrated AI systems, such as due to more frequent updates or retraining.

\end{itemize}

\subsection{Sub-Criterion 2.2: Compliance Checking}
\textit{Compliance Checking} serves as a critical extension of \textit{Systematic Oversight}, taking the maintained provenance records and operational logs and subjecting them to rigorous evaluation and audit trails. This process enables stakeholders, such as the RAI Governance Committee, to conduct in-depth reviews and audits, ensuring the AI system's adherence to RAI.


\subsubsection{\faCog\,\textbf{Process Metric 2.2.1: Auditing}}
\textbf{Context and Importance}:
Auditing is essential for affirming the ethical and legal compliance of AI systems, gaining particular significance in the rapidly advancing field of large-scale GenAI.
World-leading AI and governance experts suggest auditing as a key measure for managing risks in these advanced systems \cite{bengio2023managing}. 
Additionally, it's increasingly becoming a regulatory requirement globally, as evidenced by initiatives from 
Singapore \cite{SA_AIVerify}, EU \cite{floridi2022capai}, Australia NSW \cite{NSW}, and Canada \cite{CA023AICodeOfConduct}. Broadly, all metrics within this paper can be leveraged to conduct ethics-based auditing \cite{mokander2023operationalising}, a process that rigorously assesses an entity’s adherence to moral principles (i.e., Accountability).

\textbf{Key Considerations}:
\begin{itemize}
    \item \textbf{Diversified Auditing Strategy}: Develop a comprehensive auditing strategy that incorporates both internal and external audits \cite{novelli2023accountability, goodman2022ai, EUList2020, srinivasan2022role}. Tailor this strategy to the AI system's context, considering its risk levels, complexity, stakeholder involvement, and regulatory requirements. Ensure the independence of audits to provide unbiased evaluations \cite{OpenAIComment23, srinivasan2022role, loi2021towards, DOEPlaybook}.

    \item \textbf{Multi-Dimensional Audit Techniques} \cite{goodman2022ai}: Employ various audit techniques, including technical (focused on data and code), empirical (centered on measuring inputs and outputs), and governance-oriented (evaluating procedures and decisions).

    \item \textbf{Ethical and Legal Alignment}: Adhere strictly to established standards and ethical/legal frameworks \cite{OECD_AdvancingAIAcc, goodman2022ai, floridi2022capai}.

    \item \textbf{Regular Auditing}: Incorporate a schedule for regular auditing \cite{floridi2022capai, konigstorfer2022ai, DOEPlaybook} that includes pre-deployment \cite{floridi2022capai, MSComment23}, post-deployment \cite{OECD_AdvancingAIAcc, MSComment23}, and post-incident audits \cite{MsRAIStandard, MSComment23}.

    \item \textbf{Verifiable Audits}: Ensure that audit results are transparent and verifiable, using mechanisms \textbf{licenses and certifications} \cite{bengio2023managing, OpenAIComment23, wachter2017transparent}. This is particularly important for large-scale GenAI models in high-stakes domains.

    \item \textbf{Audit-Driven Improvements} \cite{matthews2020patterns}: Establish a structured process for incorporating audit findings into ongoing system improvements. Prioritize actions based on the findings' severity, impact, and feasibility of implementation.

\end{itemize}

\subsection{Resource \& Product Metrics -- Auditability}

\textit{Auditability} ensures AI systems are compliant with ethical and legal standards. The resources and products under this criterion facilitate thorough and effective auditing, enabling organizations to demonstrate the integrity and reliability of their AI systems.

\faLightbulbO\,Essential \textbf{\textit{Resources}} for \textit{Auditability}:

\begin{itemize}
    \item \textbf{Soft Laws}: Comprehensive audit frameworks, guidelines, and standards that provide a structured approach to auditing AI systems. These may include best practices for AI auditing and methodologies for conducting thorough evaluations. For example, the UK Information Commissioner’s Office published a guideline for conducting AI audits \cite{ICO2023AIAuditsGuide}. The capAI framework \cite{floridi2022capai} aligns with the EU AI Act, providing a template for conformity auditing of AI systems. Henrisken et al. \cite{henriksen2021situated} introduced a framework for end-to-end internal algorithmic auditing.
    
    \item \textbf{Hard Laws}: Legal and regulatory resources that provide up-to-date information on requirements relevant to AI systems. Examples include the proposed EU AI Act \cite{EUAIACT} and the US Blueprints for an AI Bill of Roghts \cite{WhiteHouse2023AIBillOfRights}.

    \item \textbf{AI Documentation Tools}: Employ various tools and frameworks for AI documentation, including data documentation (e.g., Datasheet \cite{gebru2021datasheets}, Data Readiness Report \cite{afzal2021data}), model documentation (e.g., Model Cards \cite{mitchell2019model}, Model Info Sheet \cite{kapoor2022leakage}), and system-level documentation (e.g., Reward Reports \cite{gilbert2023reward}, System Cards \cite{gursoy2022system, openai2023gpt4}, FactSheets \cite{arnold2019factsheets}, Software Bill of Materials \cite{xia2023empirical}, AI Bill of Materials \cite{xia2023trust}). These tools are crucial for ensuring comprehensive and transparent AI system documentation, aiding effective auditing.

    \item \textbf{Technical Tools}: Techniques such blockchain \cite{lo2022toward, liu2023decentralised} and knowledge graphs \cite{naja2022using, naja2021semantic} have also been explored for enabling provenance, auditing, and governance of AI systems.
\end{itemize}

\faPuzzlePiece\,Key \textbf{\textit{Products}} indicating effective implementation of \textit{Auditability} include documentations \cite{SA_AIVerify} and built-in system features:

\begin{itemize}
    \item \textbf{Provenance Records}: Detailed documentation on the data/model/system etc., resulting from the utilized AI documentation tools.
    \item \textbf{Audit Reports}: Detailed reports and documentation generated from auditing activities, including audit findings, recommendations, and action plans for rectification.
    \item \textbf{Compliance Certificates and Licenses}: Certificates or licenses could be issued upon successful completion of audits, indicating compliance with specific standards or regulations. These serve as formal recognition of the AI system's adherence to established norms.
    \item \textbf{Integrated System Features}: Built-in system functionalities such as automated logging features and version control.

\end{itemize}

\section{Results - Redressability}
\label{Sec:Redressability}
\textit{Redressability} serves as the linchpin for actionable recourse. It necessitates the creation of  formal mechanisms that facilitate stakeholder remediation for adverse impacts, not only fulfilling ethical and legal imperatives but also engender stakeholder trust.

\subsection{Redress-by-Design}
This refers to the preemptive incorporation and establishment of mechanisms for issue and error detection, management, and rectification within the AI system, which fortifies accountability by enabling timely and effective redress.

\subsubsection{\faCog\,\textbf{Process Metric 3.1.1: Incident Reporting and Response}}
\textbf{Context and Importance}:
Effective Incident Reporting and Response mechanisms are critical for mitigating adverse impacts of (Gen)AI systems \cite{srinivasan2022role}. These processes not only rectify issues but also facilitate the continuous evolution of AI systems. Their implementation is vital for maintaining stakeholder trust and adhering to ethical and legal standards.

\textbf{Key Considerations}:

\begin{itemize}

\item \textbf{Accessibility and Visibility} \cite{EUList2020}: Reporting channels must be accessible to all stakeholders, offering user-friendly options like online forms and hotlines. Prominent visibility of these channels encourages engagement and reporting.

\item \textbf{Structured Incident Management} \cite{DOEPlaybook, EUList2020}: Develop a structured process for handling reported incidents. This should include initial assessment, severity categorization, in-depth investigation, response planning, and execution of corrective actions, with provisions for progress tracking.

\item \textbf{Feedback Loop Integration} \cite{GoogleComment23, OECD_AdvancingAIAcc}: Systematically incorporate feedback from incident reports into the AI system's development and operational processes. This integration is crucial for ongoing system refinement, enhancing performance, and reducing future risks.


\end{itemize}

\subsubsection{\faCog\,\textbf{Process Metric 2.2.4: Built-in Redundancy}}

\textbf{Context and Importance}:
Built-in redundancy is a fundamental aspect of designing (Gen)AI systems to ensure their resilience and facilitate redress. It involves creating multiple layers of fallback mechanisms and alternative procedures, which are critical for the system’s fault tolerance \cite{EUList2020}. This aspect of design is indispensable for continuous operation, effective problem detection, and the rectification of errors or system failures, safeguarding against potential disruptions and upholding the integrity of AI systems.

\textbf{Key Considerations}:

\textbf{Multi-Modal Redundancy} \cite{EUList2020}: Implement redundancy across various dimensions of the (Gen)AI system, such as data storage, computational resources, and operational procedures. This ensures that the system remains functional and efficient, even in the face of component failures or external disruptions. Examples include OpenAI's global infrastructure redundancy plan \cite{OpenAI2023ImprovingLatencies}.

\subsection{Resource \& Product Metrics -- Redressability}
\textit{Redressability} underscores the importance of resources and products that facilitate prompt and effective responses to incidents, ensuring swift and effective response measures.

\faLightbulbO\,Essential \textbf{\textit{Resources}} for \textit{Redressability}:

\begin{itemize}
    \item \textbf{Redundancy Design Case Studies}: Utilize examples from existing AI systems to inform redundancy strategies. For instance, Tesla's autopilot system \cite{Tesla2023Autopilot}, which employs dual AI chips for decision consensus, serves as a practical model for effective redundancy implementation in AI designs.

    \item \textbf{Incident Management Tools}: Utilize integrated tools designed for comprehensive reporting, tracking, and managing incidents related to AI systems. These tools should facilitate efficient communication among stakeholders and responsible teams, ensuring effective responses.
\end{itemize}


\faPuzzlePiece\, Key \textbf{\textit{Products}} indicative of effective \textit{Redressability}:

\begin{itemize}
    \item \textbf{Incident and Response Documentation}: Detailed records of incidents, the responses undertaken, corrective action plans, and outcomes. Unlike ``Incident and Response Logging,'' which focuses on real-time logging, this documentation provides a post-event analysis, including a transparent record of actions, timelines, and responsible parties.
    \item \textbf{Integrated System Features}: This include built-in system functionalities that allow users to provide feedback and report incidents directly, and redundant components or functionalities as part of the system's design \cite{lupattern}.

\end{itemize}

\section{Discussion}
\subsection{Accountability and Liability}
In the context of (Gen)AI, distinguishing between liability and accountability is crucial yet complex, involving intricate legal and ethical dimensions. Liability pertains to the \textit{legal responsibility} for damages caused by AI systems, often encompassing financial reparations and associated with \textit{legal risks}. Accountability, on the other hand, relates to \textit{social expectations} and \textit{public answerability}, and is associated with both \textit{legal} and \textit{reputational risks}.

The integration of GenAI with external services and API-based interactions further complicates this landscape, introducing complex AI supply chains \cite{cobbe2023understanding}. 
For instance, training subsequent models based on outputs from upstream models can obscure the origins of data and decisions, thereby complicating the assignment of accountability and liability.
These scenarios exemplify how the scale and intricacy of GenAI systems present unique challenges that may not be fully addressed by existing legal frameworks.
In response, initiatives like the European Commission's AI Liability Directive \cite{EU_liability} aim to adapt liability rules for AI. This directive introduces concepts such as a rebuttable ``presumption of causality'' for victims and a strict liability regime for high-risk AI systems. 

\subsection{Implications}
\textbf{Practical Implications}: Our study provides a comprehensive metrics catalogue of AI accountability, synthesizing insights from both academic and grey literature. It offers practitioners and policymakers in AI a nuanced understanding of AI accountability, critical for developing robust AI governance and risk management frameworks.
The identified metrics for AI accountability are particularly valuable for organizations seeking to operationalize ethical AI principles, emphasizing not only the technical aspects but also the ethical, legal, and societal dimensions. Additionally, the inclusion of the GenAI perspective provides a timely insight into AI accountability in a rapidly evolving landscape. 
Furthermore, the practical application of these metrics in tools like web-based portals can significantly aid organizations in assessing and managing AI risks. 

\textbf{Theoretical Implications}: Our research contributes to the evolving discourse on AI governance and risk management by offering a refined conceptual framework for AI accountability. This framework emphasizes the interconnectedness of responsibility, auditability, and redressability, laying a foundation for future academic exploration in GenAI accountability.

\subsection{Limitations and Future Work}
\textbf{Limitations}: While the metrics presented in this study form a comprehensive catalogue for AI accountability, they currently primarily focus on a binary (yes/no) assessment, which may not fully capture the complexities of AI systems. The varying levels of RAI maturity across industries, coupled with the evolving legal landscape surrounding AI, especially GenAI, pose challenges to the universal applicability of these metrics. Additionally, the implementation of these process metrics necessitates concerted efforts from various stakeholders, including policymakers, AI developers, end-users, as well as the general public.

\textbf{Future Work}: Future research includes empirically validating the proposed metrics through user testing, which is an ongoing effort by the authors. Adapting the catalogue to accommodate diverse contexts, including varying stakeholders, lifecycle stages, and regional or global legislative standards, is essential to enhance its utility across different domains. Future research could also expand the scope of this study to include other major AI ethics principles. Furthermore, the development of more sophisticated metrics and scoring mechanisms, beyond the current binary (yes/no) approach, could significantly enhance the precision and applicability of our findings. This advancement would allow for a more nuanced assessment of AI accountability, taking into account the quality and effectiveness of the implemented processes.

\subsection{Threats to Validity}
\textbf{Internal Validity}: The primary threat to internal validity is the potential for subjective interpretation during thematic coding. While measures were taken to mitigate this (e.g., multiple reviewers), some degree of subjectivity is inevitable. Additionally, while our methodology was rigorous, there is a possibility that relevant works might not have been included, which could affect the comprehensiveness of our findings.

\textbf{External Validity}: The rapidly evolving nature of AI technology means that our findings may need updating as new developments emerge.

\textbf{Construct Validity}: The construct validity hinges on the appropriateness and comprehensiveness of the AI accountability metrics. While developed through a rigorous methodology, empirical validation is necessary to establish their effectiveness. Moreover, the classification into process, resource, and product metrics presents its own challenges. The identification of resource and product metrics is less explicit compared to process metrics, as they are often implicitly mentioned in the literature, which could affect the clarity and precision of our categorization.

\section{Conclusion}
In this paper, we present a comprehensive metrics catalogue for AI accountability, with a specifically tailored focus of GenAI. Our approach synthesizes insights from both academic and grey literature, resulting in a robust framework that encapsulates the critical dimensions of \textit{Responsibility}, \textit{Auditability}, and \textit{Redressability} in AI systems. This catalogue, comprising process, resource, and product metrics, serves as a practical tool for AI governance and risk assessment, particularly in the rapidly evolving domain of GenAI.

Our work contributes to the discourse on AI accountability, bridging the gap between theoretical concepts and practical applications. By integrating both technical and non-technical aspects, such as ethical, legal, and societal considerations, our findings offer a holistic view of accountable AI systems. Our work not only enriches the academic discourse but also provides actionable guidance for practitioners and fosters AI accountability.

\bibliographystyle{unsrt}
\bibliography{main}

\end{document}